\documentclass[fleqn,12pt]{article}
\usepackage{amsfonts,epsfig}
\usepackage{latexsym,amssymb}
\usepackage{times}

\newcommand{\ft}[2]{{\textstyle\frac{#1}{#2}}}
\def\tilde{\widetilde}

\def\1bar{1\hskip -.275cm -}
\def\2bar{2\hskip -.275cm -}
\def\3bar{3\hskip -.275cm -}

\newsavebox{\uuunit}

\makeatletter \@addtoreset{equation}{section} \makeatother

\def\bfone{\relax{\rm 1\kern-.35em 1}}

\def\bfone{\relax{\rm 1\kern-.35em 1}}
 
\newcommand{\nc}{\newcommand}

\newcommand{\bbR}{{\mathbb{R}}}

\newcommand{\CO}{{\cal O}}

\newcommand{\dd}{\partial}

\newcommand{\pls}{\!+\!}
\newcommand{\plss}{\!\!+\!\!}
\newcommand{\mis}{\!-\!}
\newcommand{\miss}{\!\!-\!\!}

\newcommand{\mathon}{\mathversion{bold}}
\newcommand{\mathoff}{\mathversion{normal}}

\nc{\be}{\begin{equation}} \nc{\ee}{\end{equation}}
\nc{\bea}{\begin{eqnarray}} \nc{\eea}{\end{eqnarray}}
\newcommand{\ben}{\begin{displaymath}}
\newcommand{\een}{\end{displaymath}}
\nc{\dalpha}{\dot{\alpha}} \nc{\dbeta}{\dot{\beta}}
\nc{\nn}{\nonumber} \nc{\non}{\nonumber\\}
\begin{document}

\renewcommand{\thefootnote}{\fnsymbol{footnote}}

\vspace*{0.1ex}

\bigskip\bigskip


\begin{center}
{\bf\Large AdS Duals of Matrix Strings}
\bigskip\bigskip\medskip

{\bf Jose F.~Morales$~^{1,2}$ and Henning Samtleben$~^1$ \medskip\\ }
{\em $^1$ ITP} \,and\, {\em Spinoza Institute}, 
{\em Utrecht, The Netherlands}\\
{\em $^2$ LNF-INFN}, {\em Frascati, Rome, Italy}\\
\smallskip

{\small morales@phys.uu.nl, h.samtleben@phys.uu.nl}

\end{center}

\setcounter{footnote}{0}

\begin{abstract}

We review recent work on the holographic duals of type II and
heterotic matrix string theories described by warped AdS$_3$
supergravities.  In particular, we compute the spectra of Kaluza-Klein
primaries for type I, II supergravities on warped AdS$_3\times S^7$
and match them with the primary operators in the dual two-dimensional
gauge theories.  The presence of non-trivial warp factors and dilaton
profiles requires a modification of the familiar dictionary between
masses and ``scaling'' dimensions of fields and operators. We present
these modifications for the general case of domain wall/QFT
correspondences between supergravities on warped AdS$_{d+1}\times
S^{q}$ geometries and super Yang-Mills theories with $16$
supercharges.

\end{abstract}

\renewcommand{\thefootnote}{\arabic{footnote}}


{\it Proceedings of the RTN workshop ``The quantum structure of
spacetime and the geometric nature of fundamental interactions'',
Leuven, September 2002}

\section{Introduction}

The AdS/CFT correspondence which in its original form
\cite{maldacena} relates M/string theory on anti-de Sitter spaces to
boundary conformal field theories has been rapidly extended to account
for more general M/string backgrounds and non-conformal field
theories. In~\cite{Itzhaki:1998dd}, a proposal was put forward
relating string theory on near horizon Dp-brane geometries to
$d=p\pls1$ dimensional super Yang-Mills (SYM) theories with sixteen
supercharges. Despite the earlier encouraging results
of~\cite{Boonstra:1998mp} and the obvious interest in these so called
domain wall/QFT dualities where holography is at work in the presence
of non-trivial warped geometries and dilaton profiles, these
non-conformal settings of the correspondence are still far from being
understood. In~\cite{hm}, we started a systematic study of this
physics. Here we review some of the the main ideas and results.

We focus on the case $d=2$ which is a particularly rich setting
for a domain wall/QFT correspondence. It relates string theory on
certain warped AdS$_3\times S^7$ backgrounds to two-dimensional
fundamental string or gauge theories, depending on whether we study
systems of fundamental or D-strings. Fundamental string solutions are
common to all five ten-dimensional string theories. The study of the
domain wall/QFT correspondence in this case is particularly
interesting since the string background is free of RR fields. The case
of D-strings is specific to type IIB and type I theory. The dual
gauge theories in these cases correspond to the matrix string models
\cite{Dijkgraaf:1997vv,Banks:1997it} proposed as non-perturbative
definitions of type IIA and heterotic string theories. The
correspondences under study here in principle provide a supergravity
description of this physics (see \cite{Johnson} for early discussions
on these ideas).

We refer the reader to \cite{hm} for more detailed discussions 
of the results reported here and a complete list of references.

\mathon
\section{Warped AdS$_{d+1}\times S^{q}$  geometries}
\mathoff

Let us start by extending the familiar dictionary between masses and
scaling dimensions on AdS spaces to domain wall/QFT correspondences
involving warped AdS$_{d+1}\times S^{q}$ geometries, applying the
ideas of \cite{Witten:1998qj}. More specifically, we consider spaces
described by
\bea
d\hat{s}^2&=& z^\omega \, ds^2=z^\omega \,\left[ {\ell^2\over z^2}\,
dx^\mu\, dx_\mu+\tilde{\ell}^2\, d\Omega_{q}\right]\;,
\label{metricp}
\eea
which typically arise as near horizon geometries of p-brane solutions
of low energy supergravities. Greek letters $\mu=0, 1, \dots, d$, here
refer to components along AdS$_{d+1}$ with $x^{d}\equiv z$, while
Arabic indices $m=d+1, \dots, D-1$ run over the $S^q$ sphere with
metric element $d\Omega_q$. For definiteness, let us consider
Dp-branes, such that $D=10$, $d=p+1$, $q=9-d$, and (for $d\neq 6$)
\bea
\omega &=&-{(d-4)^2\over 4\,(6-d)} \;, \qquad
\tilde{\ell}^2~=~{\ell^2\over 4}(6-d)^2 \;, \qquad
z =~ ~{2\sqrt{c_d\, g_{\rm YM}^2\, N}\over 6-d}\, r^{d-6\over 2}
\;,
\label{wDp}
\eea
\\
with a constant $c_d$, and $r$ denoting the distance from the brane
source. In addition, the supergravity solutions involve a non-trivial
flux for a rank $d\pls1$ form and (unless $d=4$) a running
dilaton. The Yang-Mills coupling constant $g_{\rm YM}$ carries
dimension of $[L]^{d-4}$. Perturbation theory is better organized in
terms the dimensionless 't Hooft parameter $\lambda_{\rm eff} \equiv
c_d\,g_{\rm YM}^2\, N\, \left({r\over
\alpha^\prime}\right)^{d-4}$. The gauge theory is weakly coupled for
$\lambda_{\rm eff}\ll 1$. The other two relevant parameters are the
AdS radius $\ell^2\, z^\omega\sim\sqrt{N}\, \lambda_{\rm
eff}^{d-4\over4}$ and the string coupling constant
$e^{\phi}\sim \lambda_{\rm eff}^{8-d\over 4}/ N$
\cite{Itzhaki:1998dd}.  As usual, genus expansion corresponds to an
expansion in ${1\over N}$. We will always work in the limit of large
$N$ with $\lambda_{\rm eff}$ kept fixed where both supergravity and
string perturbation can be trusted.

The equation of motion of a massless ten-dimensional scalar field 
moving freely on (\ref{metricp}) can be written as
\bea
\left(\square_{\rm AdS}+{4\omega\over \ell^2}\,\partial_z 
- m^2\right)\, \phi=0 \;.
\label{scalar} \eea
where $\square_{\rm AdS}$ denotes the pure AdS d'Alembertian and the
effect of the warped geometry shows up in the $\omega$-dependent shift
of the differential operator in (\ref{scalar}).  The mass parameter
$m$ is defined by the eigenvalue equation:
\bea
\square_{S^{9-d}}\,\phi=-m^2\, \phi \;,
\nn
\eea
and characterizes the harmonic mode of the scalar field $\phi$ along
the sphere $S^{9-d}$. After a Fourier transform in the $d$-dimensional
space spanned by ${\bf x}=(x^0, \dots, x^{d-1})$,
the scalar equation (\ref{scalar}) reduces to a second order
differential equation for $\phi_p(z)$ of the kind
\bea
\left(z^2\, \dd_z^2 + \left(1-2\,a \right)\, z\,\dd_z +
p^2\,z^2+a^2-\delta^2 \right)\,\phi_p(z) \;,
\label{diffeq}
\eea
with momentum $p$, and parameters
\bea
 a ~=~{d\over 2}-2\omega
\;,\qquad
\delta ~=~ \sqrt{a^2+ m^2 \ell^2} \;.
\label{cs}
\eea
Its general solution can be expressed in terms of Bessel functions
$J_\delta, Y_\delta$, as
\bea \phi_p(z) &=& z^{a}\left[ d_1 \, J_\delta (pz) +
d_2\,Y_\delta(pz)\right]
~\approx~ z^{E_0-2\,\delta} \left( \phi_{\rm def} + .... \right)
+ z^{E_0} \left( \phi_{\rm vev} + .... \right)
 \label{Bessel}
 \eea
where dots stands for subleading terms in the expansion near the
boundary $z\sim 0$. The quantities $\phi_{\rm def}, \phi_{\rm vev}$
encode the boundary conditions and represent as usual deformations and
non-trivial vev's in the dual gauge theory.  Finally $E_0$, to which
we will refer as the ``energy'', is given by
\bea E_0&=& a+\delta=a+\sqrt{a^2+m^2\, \ell^2} \;, \label{u0} \eea
In the conformal case $d=4$, the energy $E_0$ is identified with the
conformal dimension of the dual operator and governs two-point
correlation functions in the gauge theory side.  This result
generalizes straightforwardly to our case.  By plugging the solution
(\ref{Bessel}) in the equation of motion of the scalar field in the
warped geometry and following the standard holographic prescription
\cite{Freedman:1998tz} for extracting the two-point function of scalar
fields one finds~\cite{hm}
\be \langle {\cal O}({\bf x}){\cal O}({\bf
 y})\rangle\sim |{\bf x}-{\bf y}|^{-2\left(E_0-a+{d\over 2}\right)}
\;.
\label{cd}
\ee
The entire information about the warped geometry is contained in the
parameters $E_0, a$. In particular, the mass bound at which the
argument of the square root becomes negative is shifted in the warped
geometry with respect to its pure AdS cousin. A rather non-trivial
consistency check of the whole picture follows from the need of
certain conspiracies between masses and warp factors entering
$\delta$ in (\ref{u0}) in order to produce rational numbers as output
of the square root. We will verify this by explicit calculations of
the spectrum of harmonics in warped AdS$_3\times S^7$.

\mathon
\section{Type II supergravities on warped AdS$_3 \times S^7$}
\mathoff

The spectrum of primaries in type II supergravities on near horizon
D-string geometries is found by expanding the equations of motion to
linear order in the fluctuations around the corresponding warped
AdS$_3 \times S^7$ vacuum.  Ten-dimensional fluctuations are written
in terms of $S^7$-sphere harmonics $Y_{\bf m}^{\ell}(y)$ as
\be
\Phi_{\bf \mu m}=\sum_{\ell}
\phi^\ell_\mu(x)\,Y_{\bf m}^{\ell}(y)
\ee
with collective indices ${\bf \mu}, {\bf m}$ carrying the $SO(3)\times
SO(7)$ Lorentz representation of the given field.  The sum over ${\ell}$
runs over a set of allowed representations of the $SO(8)$ isometry group.
In general, the spectrum of $SO(q+1)$ representations appearing in the
Kaluza-Klein reduction of a D-dimensional supergravity on the sphere
$S^q$ is essentially determined by group theory \cite{Salam:1981xd}
(see \cite{deboer} for recent applications). The
harmonic expansion of a field transforming in the ${\cal R}_{SO(q)}$
representation of the Lorentz group of $S^q$ comprises all the
representations ${\cal R}_{SO(q+1)}$ of the isometry group $SO(q+1)$
that contain ${\cal R}_{SO(q)}$ in their decomposition.

The on-shell field content of ten-dimensional type II supergravities
$({\bf 8_v}\pls{\bf 8_s})\times({\bf 8_v}\pls{\bf 8_{s,c}})$ decomposes
under the $SO(7)$ Lorentz group of $S^7$ as
\bea
{\cal R}_{SO(7)}^{\rm II} &=&
3\cdot {\bf 1}+3\cdot {\bf 7}+4\cdot {\bf 8}+
2\cdot {\bf 21}+{\bf 27}+{\bf 35}+2\cdot {\bf 48} \;.
\label{so7a}
\eea
Every representation in (\ref{so7a}) gives rise to a tower of
harmonics of those $SO(8)$ representations in which it is contained
upon decomposition under $SO(7)$ (see table 1 of \cite{hm}).
Collecting the various contributions coming from (\ref{so7a}) and
grouping them into supermultiplets of the superalgebra of background
isometries~\cite{hm} we arrive at the spectrum of $SO(8)$
representations given by
\bea
{\cal H}_{\rm IIA}&=&
 \sum_{n=0}^{\infty} {\bf (n000)}_{\rm IIA}~\equiv~ \sum_{n=0}^{\infty}
 ({\bf 8_v} - {\bf 8_s})({\bf 8_v} - {\bf 8_s})(n000)\;,\nn\\
{\cal H}_{\rm IIB} &=& \sum_{n=0}^{\infty} {\bf (n000)}_{\rm
IIB}~\equiv~\sum_{n=0}^{\infty} ({\bf 8_v} - {\bf 8_s})({\bf 8_v} -
{\bf 8_c})(n000) \;,
\label{sumB}
\eea
for type IIA and IIB, respectively. The supermultiplets ${\bf
(n000)}_{\rm IIA,IIB}$ are displayed in
tables~\ref{specIIA},~\ref{specIIB} (the omission of any
representation with negative Dynkin labels is always understood).
\begin{table}[bt]
\centering
\begin{tabular}{||l|l|l|l|l||}
\hline \hline \footnotesize $(n\plss2000)$ & \footnotesize
$(n\plss1010)$ & \footnotesize $(n100)$ & \footnotesize $(n001)$ &
\footnotesize $(n000)$ \\
\footnotesize $(n\pls1010)$ & \footnotesize $(n100)\pls(n020)$ &
\footnotesize $(n001)\pls(n\miss1 110)$ & \footnotesize
$(n000)\pls(n\miss1 011)$  &
\footnotesize $(n\miss1 010)$ \\
\footnotesize $(n100)$ & \footnotesize $(n001)\pls(n\miss1 110)$ &
\footnotesize $(n000)\pls(n\miss1 011)\pls(n\miss2 200)$ &
\footnotesize $(n\miss1 010)\pls(n\miss2 101)$  &
\footnotesize $(n\miss2 100)$ \\
\footnotesize $(n001)$ & \footnotesize $(n000)\pls(n\miss1 011)$ &
\footnotesize $(n\miss1 010)\pls(n\miss2 101)$ & \footnotesize
$(n\miss2 100)\pls(n\miss2 002)$  &
\footnotesize $(n\miss2 001)$ \\
\footnotesize $(n000)$ & \footnotesize $(n\miss1 010)$ &
\footnotesize $(n\miss2 100)$ & \footnotesize $(n\miss2 001)$  &
\footnotesize $(n\miss2 000)$ \\
\hline \hline
\end{tabular}
\caption{\small Spectrum of IIA supergravity on $S^7$, the multiplet
${\bf (n000)}_{\rm IIA}$.}
\label{specIIA}
\end{table}
\begin{table}[bt]
\centering
\begin{tabular}{||l|l|l|l|l||}
\hline \hline \footnotesize $(n\plss2000)$ & \footnotesize
$(n\plss1001)$ & \footnotesize $(n100)$ & \footnotesize $(n010)$ &
\footnotesize $(n000)$ \\
\footnotesize $(n\pls1010)$ & \footnotesize
$(n\plss1000)\pls(n011)$ & \footnotesize $(n001)\pls(n\miss1 110)$
& \footnotesize $(n\miss1100)\pls(n\miss1 020)$  &
\footnotesize $(n\miss1 010)$ \\
\footnotesize $(n100)$ & \footnotesize $(n010)\pls(n\miss1 101)$ &
\footnotesize $(n000)\pls(n\miss1 011)\pls(n\miss2 200)$ &
\footnotesize $(n\miss1 001)\pls(n\miss2 110)$  &
\footnotesize $(n\miss2 100)$ \\
\footnotesize $(n001)$ & \footnotesize $(n\mis1100)\pls(n\miss1
002)$ & \footnotesize $(n\miss1 010)\pls(n\miss2 101)$ &
\footnotesize $(n\miss1000)\pls(n\miss2 011)$  &
\footnotesize $(n\miss2 001)$ \\
\footnotesize $(n000)$ & \footnotesize $(n\miss1 001)$ &
\footnotesize $(n\miss2 100)$ & \footnotesize $(n\miss2 010)$  &
\footnotesize $(n\miss2 000)$ \\
\hline \hline
\end{tabular}
\caption{\small Spectrum of IIB supergravity on $S^7$, the multiplet
${\bf (n000)}_{\rm IIB}$. }
\label{specIIB}
\end{table}
The quantum numbers $\ell_0, \bar{\ell}_0$, defined in terms of the
energy $E_0=\ell_0+\bar{\ell}_0$ and spin $s=\ell_0-\bar{\ell}_0$ may
be extracted from these tables as follows: The state in the upper left
corner has $(\ell_0,\bar{\ell}_0)=(\ft14(n\pls2),\ft14(n\pls2))$. The
8 unbroken supersymmetry generators act vertically in this table,
increasing the value of $\bar{\ell}_0$ from top to bottom by $\ft12$
per row. The value of $\ell_0$ is increased from left to right by
$\ft12$ per column which may be thought of as the action of the broken
supersymmetry generators. These assignments of energy and spin quantum
numbers have been verified in \cite{hm} by explicitly solving the
linearized equations motion in the near horizon string vacua.

Notice that although after reduction to $SO(7)$, the type IIA and type
IIB $SO(7)$ content (\ref{so7a}) coincides and therefore leads to the
same tower of $SO(8)$ states, the structure of the supermultiplets is
substantially different. The spectrum of type IIA on warped
AdS$_3\times S^7$ can alternatively be derived form that of
eleven-dimensional supergravity on AdS$_4\times S^7$
\cite{Biran:1983iy} upon dimensional reduction. Indeed, the states
displayed in table~\ref{specIIA} recombine into a single multiplet of
the $OSp(8|4,\bbR)$ isometry group of the eleven-dimensional vacuum.

\mathon
\section{Type I supergravity on warped AdS$_3\times S^7$}
\mathoff

The above analysis extends straightforwardly to type I supergravity on
warped AdS$_3\times S^7$. The starting point is now given by the
decomposition under $SO(7)$ of the ten-dimensional field content
$({\bf 8_v}+{\bf 8_s})({\bf 8_v}+n_v\cdot {\bf 1})$ (with $n_v$ the
number of vector multiplets):
\be
R_{SO(7)}^{\rm I}= (2\pls n_v) \cdot {\bf 1}
+(2\pls n_v)\cdot {\bf 7}+(2\pls n_v)\cdot {\bf 8}
+{\bf 21}+{\bf 27}+ {\bf 48} \;.
\label{so7I}
\ee
Again using table 1 of \cite{hm} one finds:
\bea {\cal H}_I= \sum_{n=0}^{\infty} ({\bf 8_v}+n_v\cdot {\bf 1})
 ({\bf 8_v}-{\bf 8_s})
{\bf (n000)}
\label{sumI}
\eea
The entire spectrum is collected in table~3. The quantum numbers for
states sitting in the one of the first three columns can be read off
directly from those in type IIB computed above. On the other hand, the
state on top of the last column in table~\ref{specI} carries energy
$E_0={n\over 2}+{3\over 2}$ and spin $s=1$.

\begin{table}[bt]
\centering
\begin{tabular}{||l|l|l|l||}
\hline \hline \footnotesize $(n\plss2000)$   & \footnotesize
$(n100)$   &
\footnotesize $(n000)$ &
\footnotesize $n_v \cdot(n\plss1000)$ \\
\footnotesize $(n\pls1010)$  & \footnotesize $(n001)\pls(n\miss1
110)$   &
\footnotesize $(n\miss1 010)$& \footnotesize  $n_v \cdot(n 010)$\\
\footnotesize $(n100)$  & \footnotesize $(n000)\pls(n\miss1
011)\pls(n\miss2 200)$   &
\footnotesize $(n\miss2 100)$ & \footnotesize$n_v \cdot(n\miss1 100)$\\
\footnotesize $(n001)$  & \footnotesize $(n\miss1 010)\pls(n\miss2
101)$   &
\footnotesize $(n\miss2 001)$ & \footnotesize$n_v \cdot(n\miss1 001)$\\
\footnotesize $(n000)$   & \footnotesize $(n\miss2 100)$    &
\footnotesize $(n\miss2 000)$ & \footnotesize$n_v \cdot(n\miss1 000)$\\
\hline \hline
\end{tabular}
\caption{\small Spectrum of I supergravity on $S^7$. }
\label{specI}
\end{table}

\section{Chiral primaries in the gauge theories}

We shall finally compare the spectra of KK primaries found above with
that of primary operators in the boundary theories.  The nature of the
dual theory is substantially different depending on whether we
consider systems of fundamental or D-strings. In the former case,
states in a floor of the Kaluza Klein tower below level $N$ are
associated to fundamental string states with charge $N$. The physics
at finite $g_s$ is presumably described by deformations of the
two-dimensional ${\cal N}=(8,8)$ and ${\cal N}=(8,0)$ sigma models
associated to ten-dimensional strings on flat spacetimes.  In the
D-string case the dual gauge theory is defined via quantization of the
lowest open string modes governing the low energy dynamics of $N$
nearby D-strings. They result into effective $U(N)$ and $SO(N)$
two-dimensional gauge theories for the type II and type I D-string,
respectively. We focus on these D-string systems.

Let us first consider the D-string system in type IIB. The effective
$U(N)$ gauge theory describing the low energy dynamics of $N$ nearby
D-strings is obtained from dimensional reduction of ${\cal N}=1$ SYM
in $D=10$ down to two dimensions. The field content comprises (besides
the gauge vector field) eight adjoint scalars $\phi^I$, and left and
right moving fermions $S^a$, $S^{\dot{a}}$, transforming in the ${\bf
8_v}$, ${\bf 8_s}$, and ${\bf 8_c}$, respectively, of the $SO(8)$
R-symmetry group.  The theory is manifestly invariant under sixteen
supersymmetries. The Poincar\'e symmetry group $ISO(1,1)\times SO(8)$
and supersymmetries match the isometries and Killing spinors of the
warped AdS$_3\times S^7$ background, as expected.

The analysis of primary operators follows straightforwardly from that
of ${\cal N}=4$ SYM from which the two-dimensional gauge theory
descends upon dimensional reduction. As in the four-dimensional case,
chiral primaries are associated to completely symmetrized, traceless
operators $\CO_m\equiv{\rm Tr} \left(\phi^{I_1}\dots
\phi^{I_m}\right)$, $m=2, 3, \dots$, built from scalar fields in the
${\bf 8_v}$ and transforming in the $(m000)$ representation of the
$SO(8)$ R-symmetry group.  The missing of the $m=1$ state is due the
fact that $\phi^I$ is an $SU(N)$ rather than a $U(N)$ matrix. The
remaining primaries can be found by acting with the fermionic charges
$Q^a, \tilde{Q}^{\dot{a}}$ on the chiral primary. Half of these
charges $Q^{a}$ realize the eight on-shell supersymmetries and span
the columns of table \ref{specIIB}.  The remaining
$\tilde{Q}^{\dot{a}}$ act horizontally between the various
columns. The state at the bottom right is reached by $Q^4\,\tilde{Q}^4
\CO_m= {\rm Tr}\, \left({\cal F}^4\,\phi^{I_1}\dots
\phi^{I_{m-4}}\right)$. The results agree with those coming from the
KK harmonic analysis above.

This analysis can be easily extended to the type I case.  The D-string
bound state dynamics is now described by an $O(N)$ gauge theory with
eight scalars $\phi^I$ and right moving fermions $S^a$ in the
symmetric representation of $O(N)$, eight left moving adjoint fermions
$S^{\dot{a}}$, and 32 left moving fermions $\lambda^i$ transforming as
singlets of $SO(8)$ and bifundamentals of $O(N)\times SO(32)$. The
different $O(N)$ representations of left and right moving fermions
compensate for the relative sign in the action of the
$\Omega$-projection on type IIB D-string fields. The surviving
supersymmetry is ${\cal N}=(8,0)$.  The extra fundamentals arise from
quantization of D1-D9 open strings. Chiral primaries associated to
bulk supergravity modes can be derived from those in type IIB theory
via $\Omega$-projection. From the gauge theory point of view, the
projection of operators in the second and fourth columns of
table~\ref{specIIB} can be easily understood since they are associated
to traces involving an odd number of antisymmetric matrices. In
addition, we have $n_v={\rm dim}\left[SO(32)\right]$ extra ${\cal
N}=(8,0)$ multiplets descending from the gauge chiral primary ${\rm
Tr}\, \left(\lambda^{[i} \, \lambda^{j]}\, \phi^{I_1}\dots
\phi^{I_{m-1}}\right)$ in the $(m\!-\!1000)$ representation of
$SO(8)$. This agrees with the supergravity spectrum in
table~\ref{specI}.

\bigskip

{\bf Acknowledgements:} J.F.M.  wishes to thank the organizers of the
workshop ``The quantum structure of spacetime and the geometric nature
of fundamental interactions'' in Leuven, for the organization of this
nice meeting.  This work is partly supported by EU contract
HPRN-CT-2000-00122 and HPRN-CT-2000-00131.

\end{document}